\documentclass[twocolumn,showpacs,floatfix]{revtex4}
\usepackage{graphicx}
\usepackage{epsfig} 
\begin{document}
\title{ Accuracy estimate for a relativistic Hamiltonian approach to \\
        bound-state problems in theories with asymptotic freedom }
\author{Stanis\l{}aw D. G\l{}azek}
\author{Jaros\l{}aw M\l{}ynik}
\affiliation{Institute of Theoretical Physics, Warsaw University, ul. Ho\.{z}a
69, 00-681 Warsaw, Poland}
\date{July 18, 2003}
\begin{abstract}
Accuracy of a relativistic weak-coupling expansion procedure for solving 
the Hamiltonian bound-state eigenvalue problem in theories with asymptotic 
freedom is measured using a well-known matrix model. The model is exactly 
soluble and simple enough to study the method up to sixth order in the 
expansion. The procedure is found in this case to match the precision of 
the best available benchmark method of the altered Wegner flow equation, 
reaching the accuracy of a few percent. 
\end{abstract}
\maketitle
\section{Introduction}
\label{sec:i}
This article describes a foretaste study of accuracy for a recently 
proposed Hamiltonian weak-coupling expansion procedure that in 
principle can start from a local asymptotically-free quantum field 
theory and produce sufficiently small relativistic effective Hamiltonian 
eigenvalue problems that may be soluble on a computer and yield the
wave functions of bound states in that theory. The study is needed to 
determine the prospects of reaching a reasonable accuracy in the expansion 
since the strong coupling constant rises when the renormalization group 
scale is lowered toward the scale of binding mechanism \cite{af1, af2}. 
So far, successful approaches required different formulations of the 
theory at the bound-state and high-energy scales, such as Wilson's 
lattice and Feynman's diagrammatic techniques \cite{Wilson:1976kh, 
match1, match2, match3}. They also used approximations such as the 
non-relativistic limit that applies in the case of heavy quarkonia 
\cite{charm1, charm2, charm3, charm4} and helps in constructing 
effective theories in analogy with QED \cite{CasLep, krp, nrqcd1, nrqcd2}. 

The approach discussed here is the renormalization group procedure for 
effective particles (RGEP) \cite{RGEP} that stems from the application 
of the similarity renormalization group procedure \cite{similarity} to 
the light-front Hamiltonian of QCD \cite{long}. In one and the same 
scheme, RGEP produces the asymptotically free coupling constant in 
the Hamiltonians for quarks and gluons \cite{g_QCD}, provides the 
conceptual framework for constructing the whole renormalized Poincare 
algebra in terms of the creation and annihilation operators for effective 
particles \cite{algebra}, and leads to a simple first approximation in 
the case of heavy quarkonia \cite{ho}. The procedure is boost invariant 
and raises a hope for connecting the constituent model for hadrons at 
rest \cite{Hagiwara} and the parton picture in the infinite momentum 
frame \cite{partons}. The key ingredient of the procedure is the vertex 
form factor $f$ that multiplies all interaction terms. It falls off 
quickly to zero when the change of an invariant mass in an interaction 
vertex exceeds the width parameter $\lambda$. This width variable is
also the renormalization-group evolution parameter. The effective 
particles that correspond to a small value of $\lambda$ cannot be 
copiously created because the form factor makes the interactions 
effectively weak even if the coupling constant becomes large. The 
internal structure of the effective particles at such small $\lambda$ 
may still be given by the parton-like picture in terms of constituents
that correspond to a much larger value of $\lambda \sim Q$ in the same 
flow, where $Q$ is the momentum scale of the external probe \cite{largep}.

Qualitative accuracy studies of a similarity scheme were performed before
\cite{modelafbs} using the elegant Wegner flow equation \cite{Wegner1, 
Wegner2}, which was invented independently for solving Hamiltonian eigenvalue 
problems in solid state physics. Unfortunately, it was found that the Wegner 
equation was not suitable for a straightforward weak-coupling expansion 
beyond second order. The coefficients of the expansion grew too fast and 
alternated in sign, which lead to erratic results for bound state energies
with no signs of convergence. But the Wegner equation can be modified within 
the similarity scheme \cite{optimization} and the improved equation provides 
the benchmark here for estimating the accuracy prospects for the relativistic 
RGEP procedure. The comparison with the altered Wegner equation and the fact that 
improvements are also needed in condensed matter physics \cite{Mielke1, Mielke2, 
Mielke3}, imply that the effective particle approach may also find application 
outside QCD, i.e., wherever the dynamical couplings increase in the flow of 
Hamiltonians toward the region of physical interest.

\begin{figure}
\includegraphics[scale=1]{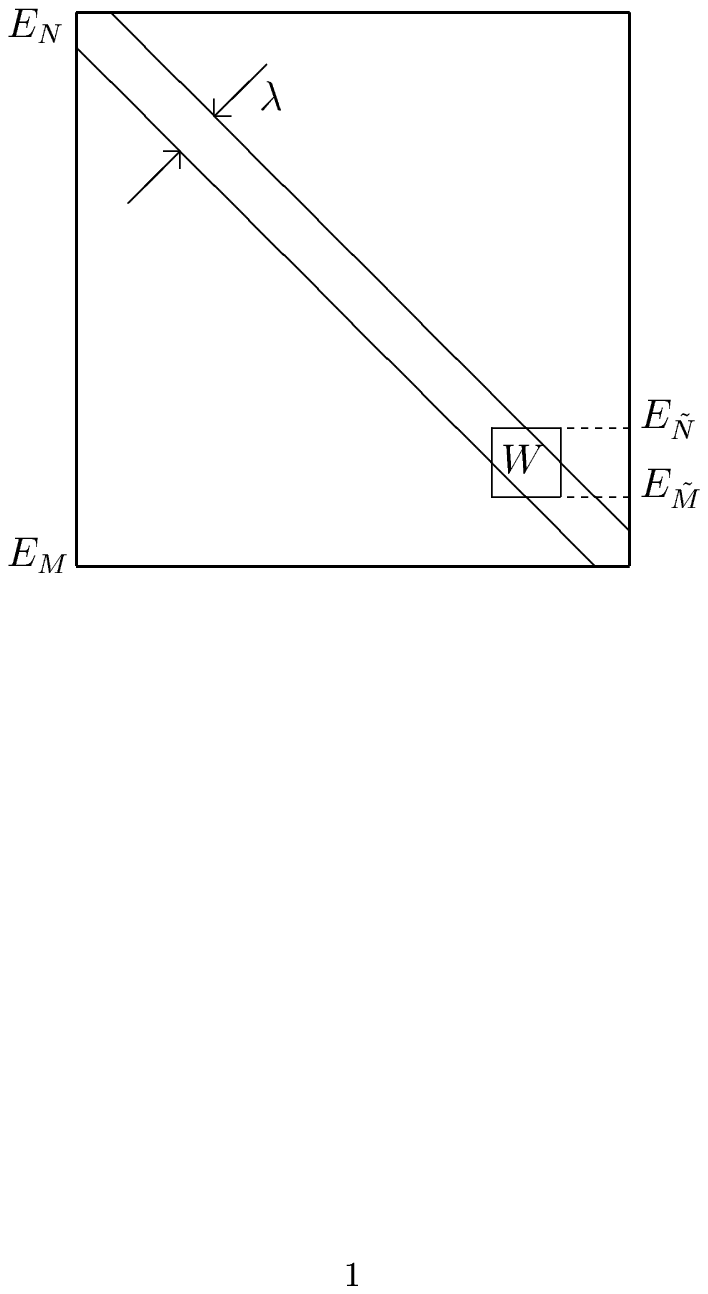}
\caption{This out-of-scale figure illustrates Eq. (\ref{W}) for the matrix 
$W_{\lambda \, mn}$ with indices $\tilde M \le m,n \le \tilde N$. The matrix 
of $H_\lambda$ is represented by the large square. Matrix elements of 
$H_\lambda$ outside the diagonal band of width $\lambda$ are equivalent to
zero. In the model study: $E_N \sim 65000$ GeV, $E_{\tilde N} \sim 4$ GeV, 
and the width $\lambda \sim 2$ GeV.}
\label{matrix}
\end{figure}

The RGEP strategy that is tested here is to start with a regulated 
$H$ of the theory to be solved (this $H$ provides an initial condition for 
the differential equations of RGEP at $\lambda=\infty$), and to evaluate 
$H_\lambda$ with $\lambda$ on the order of a bound-state energy as a series 
in powers of a coupling constant $g_\lambda$. After evaluation of $H_\lambda$, 
one calculates its matrix elements in the basis defined by eigenstates of the 
Hamiltonian $H_{0\lambda}$, which is obtaind from $H_\lambda$ by setting 
$g_\lambda=0$; $H_{0\lambda}|n\rangle = E_n |n\rangle$ for all values of the 
label $n$. Suppose that the labels are ordered so that $E_m < E_n$ implies 
$m < n$ and the initial Hamiltonian is regulated by forcing matrix elements 
$H_{mn}=\langle m|H| n \rangle$ to vanish unless $M \le m,n \le N$ with 
certain ultraviolet cutoff number $N$ and infrared cutoff $M$. The RGEP 
procedure is designed in such a way that the matrix elements $\langle m | 
H_\lambda | n \rangle$ quickly tend to 0 when $|E_m - E_n|$ grows above 
$\lambda$, see Fig. \ref{matrix}. The next step is to focus attention on the 
window $W_\lambda$ of matrix elements of $H_\lambda$ among the basis states 
that have energies similar to the energy scale $E$ of the physical problem 
at hand, i.e., 
\begin{eqnarray}
\label{W}
W_{\lambda \, m n} = \langle m|H_\lambda | n \rangle \, ,
\end{eqnarray}
with $\tilde M \le m,n \le \tilde N$ and $E_{\tilde M} \lesssim E \lesssim 
E_{\tilde N}$. Since only states within the width $\lambda$ on the energy 
scale can directly interact with each other, states that differ in energy 
from $E$ by much more than $\lambda$ are usually not important \cite{modelafbs, 
optimization}. They can be important as long as the coupling strength can
overcome the difference in energy, but it does not matter here because the 
coupling constant is assumed to not grow to large values. The next step is 
to solve the non-perturbative eigenvalue equation for the matrix $W_{\lambda 
\, m n}$ by diagonalizing it on a computer. The middle-size eigenvalues of 
$W$ are expected to be close to the exact solutions with accuracy that depends 
on many factors in the procedure. These dependencies need to be estimated. 
The main question addressed here concerns the accuracy of the weak-coupling 
expansion for $W_\lambda$.

An exact RGEP procedure provides $W_{\lambda \, mn}$ whose structure depends 
on $\lambda$ but the eigenvalues of the middle size in the window spectrum 
do not. Becasue the couplings to the states outside the window are ignored, 
the eigenvalues of sizes at the limits of the window spectrum cannot be accurate 
even if the window is calculated exactly. Now, when one calculates $W_\lambda$ 
in the expansion in powers of $g_\lambda$ to some order, and extrapolates the 
result to $g_\lambda \sim 1$, considerable errors can ensue because of the 
missing terms in the series. Moreover, one never knows the right value of 
$g_\lambda$ at given $\lambda$ from the theory. Therefore, one must fit 
$g_\lambda$ to the bound-state observables and perform consistency checks 
for a whole set of them. The smaller is $\lambda$ the larger is $g_\lambda$ 
and the more significant are the errors of perturbation theory in evaluation 
of $W_\lambda$. But at the same time the larger is $\lambda$, the larger must 
be the range $[\tilde M, \tilde N]$ of basis states needed in the computer 
diagonalization of $W_\lambda$. A compromise must be made and the critical 
question is how close one can come to a true solution using RGEP equations. 
The question is essential for the prospects of applying RGEP to QCD. A 
well-known asymptotically free matrix model is used here to find out what 
level of accuracy can in principle be expected. One should stress at this
point that even if the test gave a promising result in the model, the utility 
of the procedure would remain only tentative until the actual calculations in 
realistic theories are performed and display signs of stability as functions 
of the order of the expansion, size of $\lambda$, variation of the window,
and other more specific features of the eigenvalue problems at hand.

Section \ref{sec:model} defines the test model and quotes equations used 
for calculating $H_\lambda$ in the benchmark of the altered Wegner's flow 
and in the RGEP case. Section \ref{sec:ae} describes results obtained using 
six different ways of fitting the coupling constant to the exact spectrum. 
These ways are labeled throughout the paper by letters A, B, C, D, E, and F. 
Each of these versions is studied in six successive orders of the weak-coupling
expansion. Section \ref{sec:c} provides a brief conclusion. Appendix \ref
{app:4th} contains the generic analytic RGEP formula for 4th order calculations 
that applies to arbitrary $H$, and Appendix \ref{app:nd} provides the numbers 
that illustrate in detail what happens in the expansions including from one 
to six orders.

\section{Model}
\label{sec:model}
The matrix elements of the model Hamiltonian used here for estimating 
the accuracy of RGEP are \cite{modelafbs,optimization} 
\begin{equation}
H_{mn} \, = \, E_n \, \delta _{mn} \, - \, g \,\sqrt{E_m E_n } \, ,
\label{eq:Hmodel}
\end{equation}
where $E_n=b^n$, $b > 1$, and $n$ is an integer, with a convention that 
the energy equal 1 corresponds to 1 GeV. The model diverges and needs to 
be regulated by the ultraviolet cutoff $\Lambda = b^N$ that limits the
allowed energies $E_n$ from above. Also a low energy cutoff, $b^M$, is 
introduced for making the Hamiltonian matrix finite and enable exact 
computations, but the lower bound is of no physical consequence for the 
results reported here. Thus, the subscripts in Eq. (\ref{eq:Hmodel}) 
are limited to the range $[M,N]$, $M$ being large negative and $N$ large 
positive. The ultraviolet renormalizability of the model, its asymptotic 
freedom, and its lack of sensitivity to the infrared cutoff, were described 
in \cite{modelafbs, optimization}. 

Two values of the cutoff $N$ and the corresponding coupling constant 
$g=g_N$ are used in this study of RGEP: $g_{16} = 0.060600631$ and  $g_{20} = 
0.04878048667$. The two cutoffs are introduced to verify the accuracy of  
renormalizability in the RGEP scheme. It is known that for such large 
values of $N$ the effective dynamics with $\lambda \sim 1$ is practically 
independent of $N$ in the case of altered Wegner's equation, and one can 
verify how well the RGEP approach satisfies this condition. At the same 
time, this condition puts constraints on the range of changes that one 
can consider in varying the RGEP generator without interference with 
renormalizability (see below). The coupling constants $g_{16}$ amd $g_{20}$ 
are fitted to obtain the exact bound-state eigenvalue $E = -1$ with 8 
digits of accuracy for $b=2$ and $M=-21$, as in \cite{optimization}. With 
these choices, the Hamiltonian $H$ is a $(N-M+1)$ $\times$ $(N-M+1)$ matrix 
with $N-M$ positive eigenvalues, and one negative, spanning the range of 
energies between $E_M \sim 0.5$ KeV and $E_N \sim 65$ or 1000 TeV. 

The similarity renormalization group procedure for Hamiltonians that 
leads to the altered Wegner equation \cite{Wegner1, optimization} and 
provides the benchmark here, can be written in the differentail form as
\begin{equation}
\frac{d\mathcal{H}}{d\lambda } =
\left[ F\{\mathcal{H}\}, \mathcal{H} \right]\, ,
\label{eq:rgrwegner}
\end{equation}
with the initial condition $\mathcal{H}=H$ when $\lambda=\infty$. The 
initial condition contains counterterms but the similarity analysis showed 
\cite{similarity} that the structure of the counterterms is simple and the 
presence of them is equivalent for large $N$ to making $g$ in $H$ depend 
on $N$. This is precisely what is done by the fitting mentioned earlier that 
guarantees that the eigenvalue $E=-1$ stays unchanged for different $N$s. 
All small eigenvalues are then independent of $N$. The generator of the 
similarity transformation can be written as ($\mathcal{D}_m=\mathcal{H}_{mm}$)
\begin{equation}
\langle m| F\{\mathcal{H}\}|n \rangle =
h_{mn}(\mathcal{D}_{m}-\mathcal{D}_{n})\mathcal{H}_{mn}\, .
\label{eq:rgrmlynik}
\end{equation}
Different choices of $h_{mn}$ lead to different matrix elements of the
renormalized Hamiltonians. Assuming
\begin{equation}
h_{mn}=\phi _{mn}\, {ds \over d\lambda} \, ,
\label{eq:phi}
\end{equation}
one obtains Wegner's equation when $\phi _{mn} \equiv 1$, and
$s=1/\lambda^2$ plays the role of the original Wegner parameter 
{\it l} \cite{Wegner1}. The altered Wegner equation has \cite{optimization}
\begin{eqnarray}
\label{factorphi}
\phi _{mn} = {1 \over 1 + |m-n| } \, ,
\end{eqnarray}
and this new equation is referred to as the benchmark.

In the plain perturbative RGEP procedure, the matrix elements of 
$H_\lambda$ in the effective basis states associated with the scale 
$\lambda$, are obtained from the matrix elements of an auxiliary 
Hamiltonian $\mathcal{H}$ in the initial basis \cite{RGEP}. The 
structure of $\mathcal{H}$ is given by
\begin{eqnarray}
\mathcal{H} \, = \, \mathcal{H}_0 \, + \, f \mathcal{G}_I\, .
\label{eq:deffiG}
\end{eqnarray}
The Hamiltonian $\mathcal{H}_0$ is equal to the initial $H_0$, which is
the free part of $H$, i.e., $H$ with $g=0$. $f$ denotes the form factor that 
can be written in all matrix elements using eigenvalues of $\mathcal{H}_0$, 
and reads
\begin{eqnarray}
f_{mn}=\exp{ \left[ - \, \phi _{mn} \, \frac{(E_m-E_n)^2}{\lambda^2}\right] }\, .
\label{eq:deff}
\end{eqnarray}
The RGEP equation for $\mathcal{G}_I$ is
\begin{equation}
\label{master}
\frac{d\mathcal{G}_I}{d\lambda }
=\left [f\mathcal{G}_I,\, \left\{ {d\over d\lambda}\, 
(1-f)\mathcal{G}_I \right\} \right ]\, ,
\label{eq:rgrglazek}
\end{equation}
where the curly bracket around an operator has the following 
meaning in terms of the matrix elements,
\begin{equation}
\left\{ A \right\}_{mn}=\frac{A_{mn}}{E_n-E_m}\, .
\label{eq:defnawklamr}
\end{equation}

A priori, the optimal choice for $\phi_{mn}$ in the RGEP could be 
different from the one that optimized the benchmark \cite{optimization}. 
But we have studied various factors $\phi_{mn}$ a la \cite{optimization} 
and found that $c \sim 1$ is also the best choice to make in RGEP, for 
similar reasons. In addition, it is useful for the test that these factors 
are made equal since then the first order calculations give identical 
results in both approaches. The factor $\phi_{mn}$ is included in the 
analytic 4th order RGEP formulae provided in Appendix \ref{app:4th}.

The RGEP Eq. (\ref{master}) cannot be integrated exactly on a computer 
as easily as Wegner's equation can, because it contains the derivative 
of $\mathcal{G}_I$ on its right-hand side. One has to solve a complex 
linear problem to extract ${\mathcal G}'_I$. But the altered Wegner 
equation provides a perturbative benchmark pattern that is known to 
approximate an exact solution well and one can estimate the accuracy 
of RGEP by comparison. 

In the perturbative evaluation of $H_\lambda$, the RGEP calculus 
is free from the problem of extracting ${\mathcal G}'_I$ because 
the derivative is computed order by order and all terms needed on 
the right-hand side are known at each successive order from the 
lower order results. In fact, the new procedure is designed for
a perturbative approach. It is simpler than the Wegner case in the 
sense that the form factors guarantee the band structure in perturbation
theory so that this structure does not have to be recovered from and 
controlled in the evolution of specific matrix elements. Small energy 
denominators that might otherwise lead to infrared singularities are 
excluded by design of RGEP. It also does not generate any terms 
with inverse powers of $\lambda$ in the coefficients of products of 
the interaction Hamiltonian (see Appendix \ref{app:4th} and Eq. (3.2)
in \cite{optimization}). Such terms can in principle lead to a variety 
of mixing effects in the evolution of $H_\lambda$, when one uses the 
Wegner equation. On top of these purely quantum mechanical features, 
the perturbative RGEP calculus is capable of respecting seven Poincare 
symmetries in an economic way and has a potential to obtain the remaining 
three \cite{algebra}. At the same time, it preserves the cluster 
decomposition property \cite{Weinberg} in the effective interactions. 
All these features are desired for the description of relativistic 
particles using theories with asymptotic freedom. But the accuracy 
of the weak-coupling expansion for windows $W_\lambda$ in RGEP must 
be measured against the benchmark to estimate the cost of the apparent 
advantages in terms of precision.
\section{Accuracy of RGEP}
\label{sec:ae}
The results of weak-coupling expansions up to the first-order terms
in the RGEP and altered Wegner equation are identical, and read
\begin{eqnarray}
\label{eq:1rzad}
\mathcal{H}_{mn} & = & E_m \delta _{mn} - g_\lambda \, \sqrt{ E_m E_n } \, 
f_{mn} \, .
\end{eqnarray}
The form factor $f_{mn}$ causes that the interaction Hamiltonian
matrix is narrow on the energy scale and has width $\lambda $, see 
Figure \ref{matrix}. It is clear that in the case of Eq. (\ref{eq:1rzad}) 
the effective coupling constant can be extracted from the matrix 
$\mathcal{H}$ using the formula
\begin{equation}
g_{\lambda }=1-\frac{\mathcal{H}_{M,M}(\lambda )}{E_{M}}\, ,
\label{defglambda}
\end{equation}
which is analogous to the Thomson limit in QED when $M$ is large and negative.
The same Eq. (\ref{defglambda}) is used for defining $g_\lambda$ also in higher 
order calculations. 

In the weak-coupling expansion in powers of $g_\lambda$ to order $k$
(see e.g.  \cite{optimization}), 
\begin{eqnarray}
\mathcal{H} = \mathcal{H}_0
            +g_\lambda \mathcal{H}_1
            +g_\lambda^2 \mathcal{H}_2
            +...
            +g_\lambda^k\mathcal{H}_k \, .
\label{eq:hameff}
\end{eqnarray}
Appendix \ref{app:4th} contains analytic expressions for $\mathcal{H}_k$
with $k = 1$, 2, 3, and 4. The right value of $g_\lambda$ can be found by 
solving the flow equations with the initial condition $g_\infty = g_N$ exactly. 
Then, one can check the accuracy of a perturbatively computed $H_\lambda$ 
by diagonalizing it for the exact value of $g_\lambda$, and by comparing the 
resulting bound-state eigenvalue with $E=-1$ that was secured to exist by 
the initial choice of $g_N$ \cite{modelafbs, optimization}. The accuracy 
test for RGEP is carried out differently because the exact solution of 
Eq. (\ref{master}) is not known. This situation is analogous to QCD where 
one can use perturbation theory to calculate $H_\lambda$ but an exact value 
of $g_\lambda$ is not available as a function of $g_N$. For the purpose 
of the accuracy test, the exact spectrum of the model is 
treated as experimental data. An approximate value of $g_\lambda$ is found 
by fitting some eigenvalue of the perturbatively calculated $W_\lambda$ to 
the data, or by performing a fit for a whole group of eigenvalues. Then one 
checks how well the bound-state eigenvalue $E=-1$ is reproduced using the 
best-fit value for $g_\lambda$. In principle, one could fit $g_\lambda$ at 
one value of $\lambda$ that is most convenient for that purpose, evolve this 
value using RGEP to the new $\lambda$ that is most suitable for the bound-state 
calculation, and then compare the calculated spectrum with data \cite{g1, match3, 
g2, g3}. In fact, the model used here can be used for testing accuracy of 
such procedures in a comprehensive way. This type of tests may help 
in narrowing the current spread of estimates for $\alpha_s$ that come from 
various sources \cite{Hagiwara}, by distinguishing theoretical procedures
of least ambiguity. But the accuracy of RGEP is checked here using one and 
the same scale for fitting $g_\lambda$ and calculating the bound-state 
energy, for simplicity. The scale we choose is $\lambda = 2$.

There exist infinitely many options for how one can fit $g_\lambda$ so 
that the spectrum of the window $W_\lambda$ at $\lambda = 2$ approximates 
the exact one as closely as possible. We display results for six options 
that illustrate the dependence of results on such choices, labeled by A, 
B, C, D, E, and F. All methods are based on the minimalization of certain 
function $K(g_\lambda)$. We use
\begin{equation}
K_r(g_\lambda)= Z \sum_n \left(\frac{v_n}{v_{e \, n}}-1\right)^2 \, ,
\label{eq:fit1}
\end{equation}
and
\begin{equation}
K_s(g_\lambda)= Z \sum_n \left(\frac{v_n-v_{n+1}}{v_{e \, n}-
v_{e \, n+1}}-1\right)^{2}\, ,
\label{eq:fit2}
\end{equation}
where $v_{e \, n}$ is the exact eigenvalue of number $n$, and $v_n$ is the 
corresponding eigenvalue of $W_\lambda$ when $W_\lambda$ is derived in 
a given order $k$. $Z$ is the normalization constant equal to the inverse 
of the number of terms in the sum. $Z$ is not important in the minimalization 
of $K(g_\lambda)$ as function of $g_\lambda$. The subscript $r$ refers to 
the ratios of eigenvalues used in $K_r$, and the subscript $s$ refers to 
the splittings between the eigenvalues used in $K_s$. The six choices differ
by which function $K(g_\lambda)$ is used and what is the range of summation 
over $n$ in Eqs. (\ref{eq:fit1}) or (\ref{eq:fit2}). 

The eigenvalues of $H$ are numbered from $M$ to $N$ in the order in which 
they appear on the diagonal of $H_\lambda$ when $\lambda \rightarrow 0$ 
in the benchmark calculation. This numbering is also applied to the 
corresponding rows and columns of the matrix $H_{\lambda \, m n}$. The 
window $W_\lambda$ is always chosen to extend from $\tilde M = -8$ to 
$\tilde N = 2$ \cite{optimization}, and it has 11 eigenvalues. These are 
numbered in the same order as for $H_\lambda$, with the bound-state eigenvalue 
having number $n_0$. We distinguish two eigenvalues that are closest in modulus 
to the bound-state energy $E=-1$, one just smaller than 1, with number $n_s$, 
and one just larger than 1 with number $n_l$. The six fitting procedures are 
designated as follows:
\begin{eqnarray}
\label{caseA}
{\rm A} & \Rightarrow &  K_r , \,  n=n_s \, ,  \\
\label{caseB}
{\rm B} & \Rightarrow &  K_r , \,  n=n_l \, ,  \\
\label{caseC}
{\rm C} & \Rightarrow &  K_r , \,  n \in \{n_s,n_l\} \, ,  \\
\label{caseD}
{\rm D} & \Rightarrow &  K_s , \,  n=n_l \, ,  \\
\label{caseE}
{\rm E} & \Rightarrow &  K_r , \,  n \in [\tilde{M}+2, \tilde{N}-2] \, , n \neq n_0 \, , \\
\label{caseF}
{\rm F} & \Rightarrow &  K_s , \,  n \in [\tilde{M}+2, \tilde{N}-2] \, , n \neq n_0 \, .  
\end{eqnarray}
In cases E and F, the two smallest and two largest eigenvalues are 
dropped because they are too much distorted by the edge of $W_\lambda$, 
as explained in Section \ref{sec:i}. The description of results obtained 
in the benchmark and tested approach in six successive orders of perturbation 
theory in each of these fitting schemes involves 72 results. The benchmark 
results are labeled "Wegner" and the tested case is labeled "RGEP." Appendix 
\ref{app:nd} contains all pertinent numbers. An example is given below to 
help in reading the figures and tables in the appendix. Otherwise, only 
main features of the results are explained.

\begin{figure}
\includegraphics[scale=0.38]{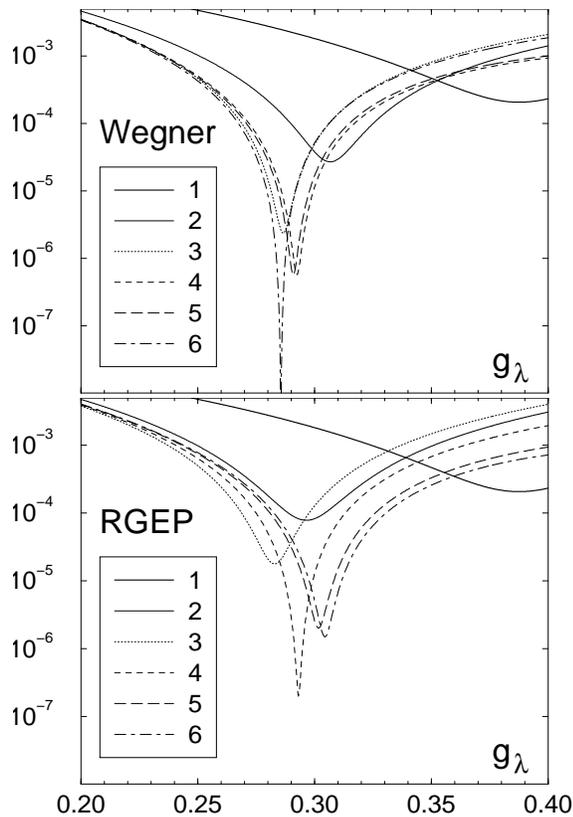}
\caption{Plots of $K(g_\lambda)$ from Eq. (\ref{eq:fit1}), in the 
altered Wegner case (benchmark), and in the RGEP procedure, as marked, 
in the fit C of Eq. (\ref{caseC}) and six successive orders of perturbation 
theory used in evaluation of the window $W_\lambda$, with the 
renormalization group parameter $\lambda = 2$ GeV, as functions of 
the a priori unknown value of the coupling constant $g_\lambda$ at 
this $\lambda$. $K(g_\lambda)$ measures deviation of two selected 
eigenvalues of the window $W_\lambda$ (those which are nearst-in-size 
to the bound-state eigenvalue) from their exact counterparts. The 
shapes of the functions point toward the required values of $g_\lambda$. 
This example is provided for explanation of how to read other examples 
given in Appendix \ref{app:nd}.}
\label{fig:fitprz}
\end{figure}

Fig. \ref{fig:fitprz} corresponds to the case C of Eq. (\ref{caseC}). 
The shape of $K(g_\lambda)$ clearly selects the value of $g_{\lambda }$ 
preferred by a given fitting procedure. Similar plots in other cases are 
given in Fig. \ref{fitowanie} in Appendix \ref{app:nd}. The numbers
that result from the fits for the coupling constants, along with the 
corresponding bound-state eigenvalues of $W_\lambda$, are also tabulated 
in Appendix \ref{app:nd}. 

\begin{figure}
\includegraphics[scale=0.3,angle=270]{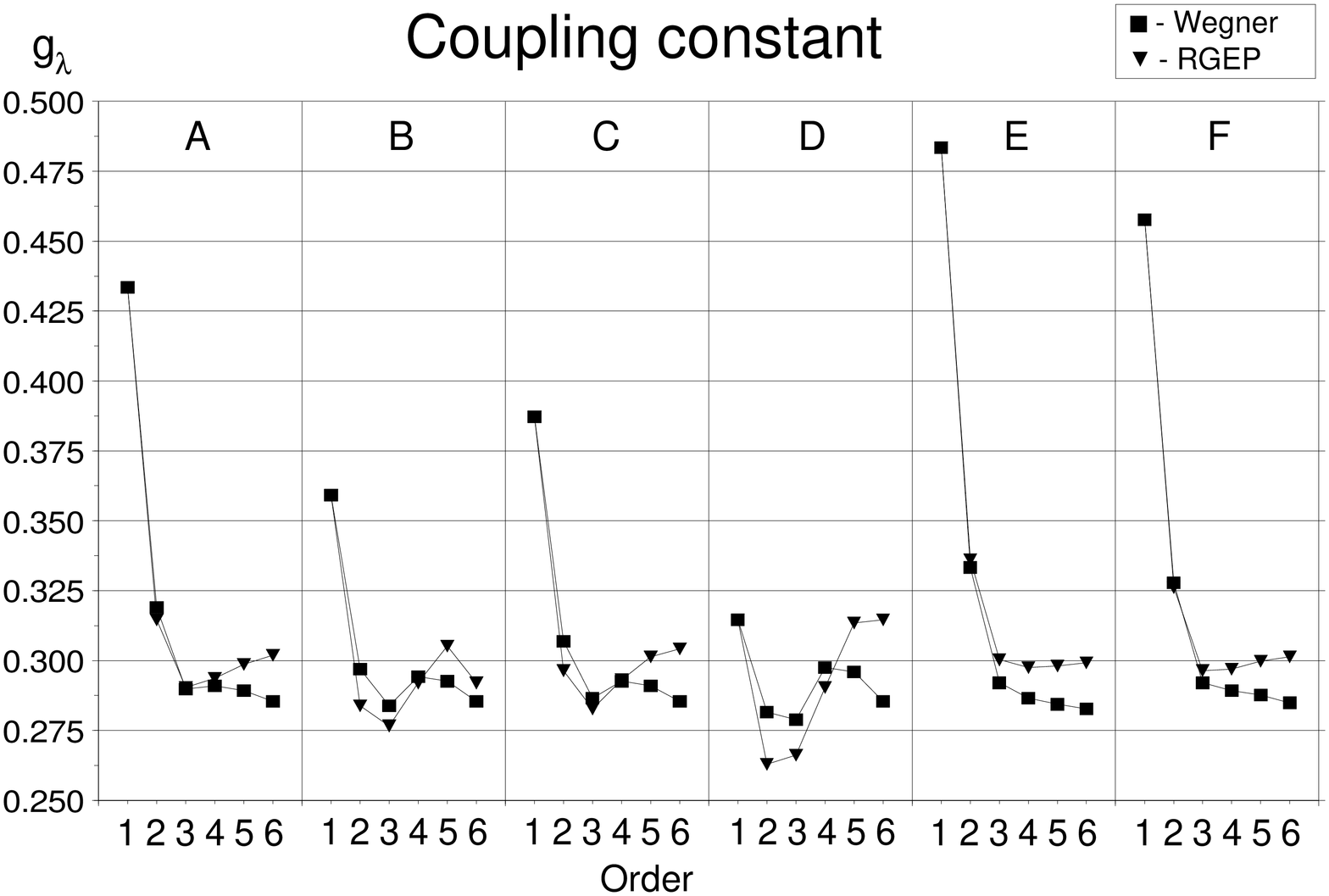}
\caption{Coupling constants $g_\lambda$ obtained from fits described 
in Eqs. (\ref{caseA}) to (\ref{caseF}), gathered in the columns 
labeled by A to F, in the benchmark case (Wegner) and the tested case 
(RGEP). The six entries in each column correspond to six successive 
orders of perturbation theory used in the calculation of the window 
$W_\lambda$ with $\lambda = 2$ GeV.}
\label{glambda}
\end{figure}

The summary of results for the coupling constants obtained in the fits 
is given in Fig. \ref{glambda}. The numbers 1 to 6 on the horizontal 
axis correspond to the order of perturbation theory in the evaluation 
of $W_\lambda$. The columns labeled A to F correspond to the algorithms
given in Eqs. (\ref{caseA}) to (\ref{caseF}). The RGEP calculation equals
the benchmark in the first order results. It is also visible that the 
fits of $g_\lambda$ in the benchmark case consistently reproduce the 
exact value of $g_\lambda = 0.2852$ at $\lambda = 2$ GeV \cite{optimization}. 
The RGEP 
displays similar stability and coalesces around 0.3. However, when the 
nearst neighbor level with energy larger than 1 is included in a fit as 
the only one (case B), or together with a nearest lower level using ratios 
of eigenvalues (case C), or together with the nearest lower level but 
using ratios of splittings between the two eigenvalues (case D), a visible 
variation in the fits occurs. This variation can be attributed to the 
lack of accuracy in the calculation of the nearst higher level, since 
in the cases E and F that include additional five lower levels, the 
higher level becoming much less significant, the fits resemble case A 
where the higher level is absent. This result suggests a rule for fits 
in future calculations that they should be focused on states with eigenvalues 
as far as possible from the bounds implied by the window choice and 
$\lambda$, in order to avoid the lack of convergence. Note that even in 
the benchmark case the 4th order calculation has to be corrected in the 
orders 5th and 6th to bring the accuracy into the few percent range 
around the exact value of $g_\lambda$. The same effect is observed in 
the bound-state eigenvalues themselves.

\begin{figure}
\includegraphics[scale=0.3,angle=270]{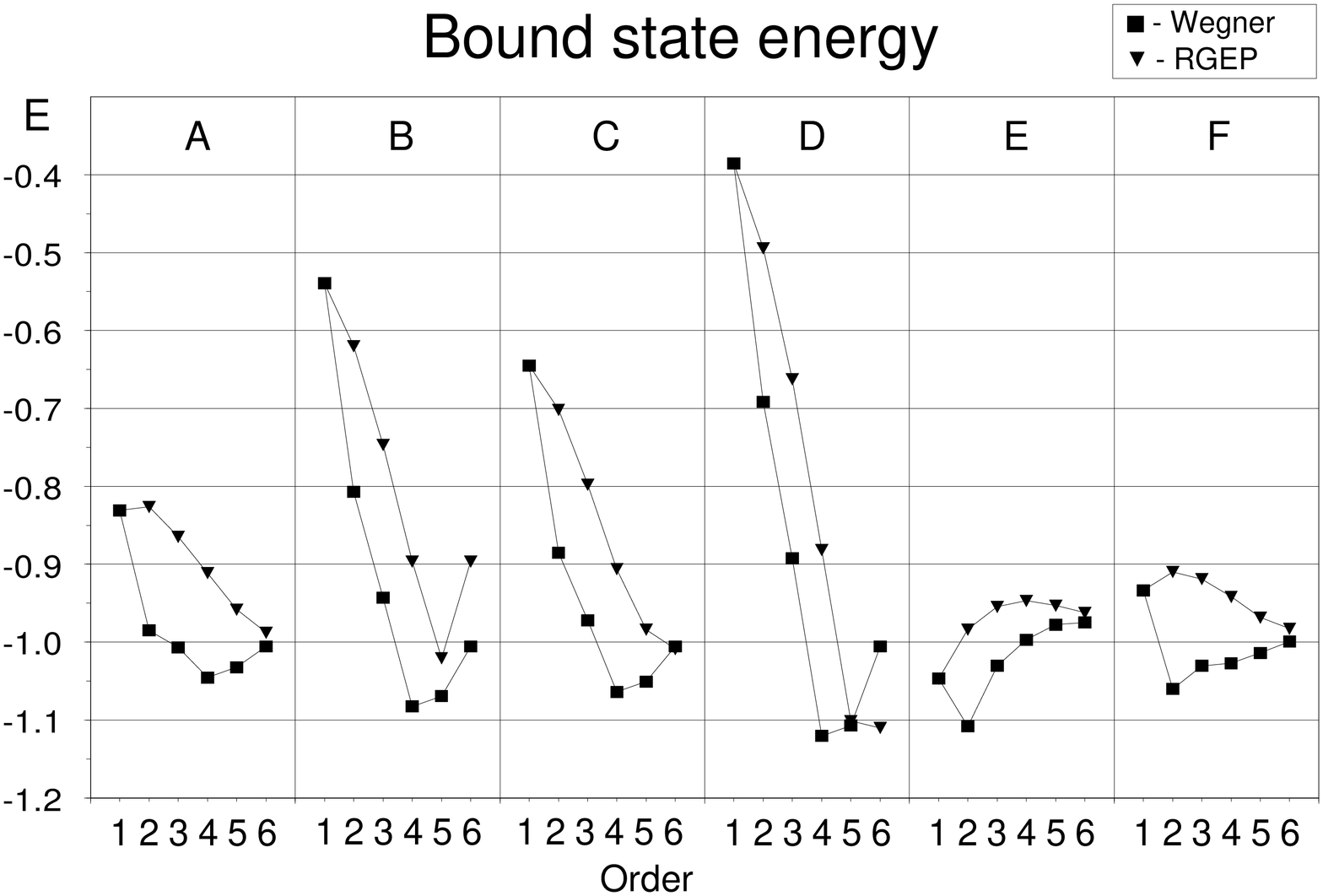}
\caption{Energies of bound states obtained from a non-perturbative 
diagonalization of window matrices $W_\lambda$ with $\lambda =2$ 
GeV. The matrices were derived in perturbation theory and evaluated 
using the corresponding values of the renormalized coupling constants, 
$g_\lambda$, from Fig. \ref{glambda}. The eigenvalues are displayed 
in the same convention and in the one-to-one correspondence to Fig. 
\ref{glambda}. Numerical values are tabulated in Appendix \ref{app:nd}. 
The exact result is -1.}
\label{boundstate}
\end{figure}

The summary of results for the bound-state eigenvalues is provided
in Fig. \ref{boundstate} in a one-to-one correspondence to Fig. 
\ref{glambda}. The eigenvalues are obtained by diagonalization of 
windows $W_\lambda$ that are calculated in the six consecutive 
orders of perturbation theory (indicated on the horizontal axis
in the individual columns), using the corresponding values of 
$g_\lambda$ from Fig. \ref{glambda} (the numbers are given in Appendix 
\ref{app:nd}). It is visible that the RGEP procedure matches accuracy 
of the altered Wegner equation. One also sees the dramatic effect of the 
attempts to include the next higher level. The results clearly point 
out that both procedures should not be used for calculations of energy 
levels close to the size of $\lambda$. The fourth order calculations
achieve accuracy on the order of 5\%. This is encouraging, because
orders 5 and 6 lead to even better results and one can expect that
fits for $g_\lambda$ can be improved by focusing on the properties 
of low energy levels, including properties other than just eigenvalues. 
It is hard to imagine that such focus could lead to worsening of the 
accuracy.

\begin{figure}
\includegraphics[scale=0.4]{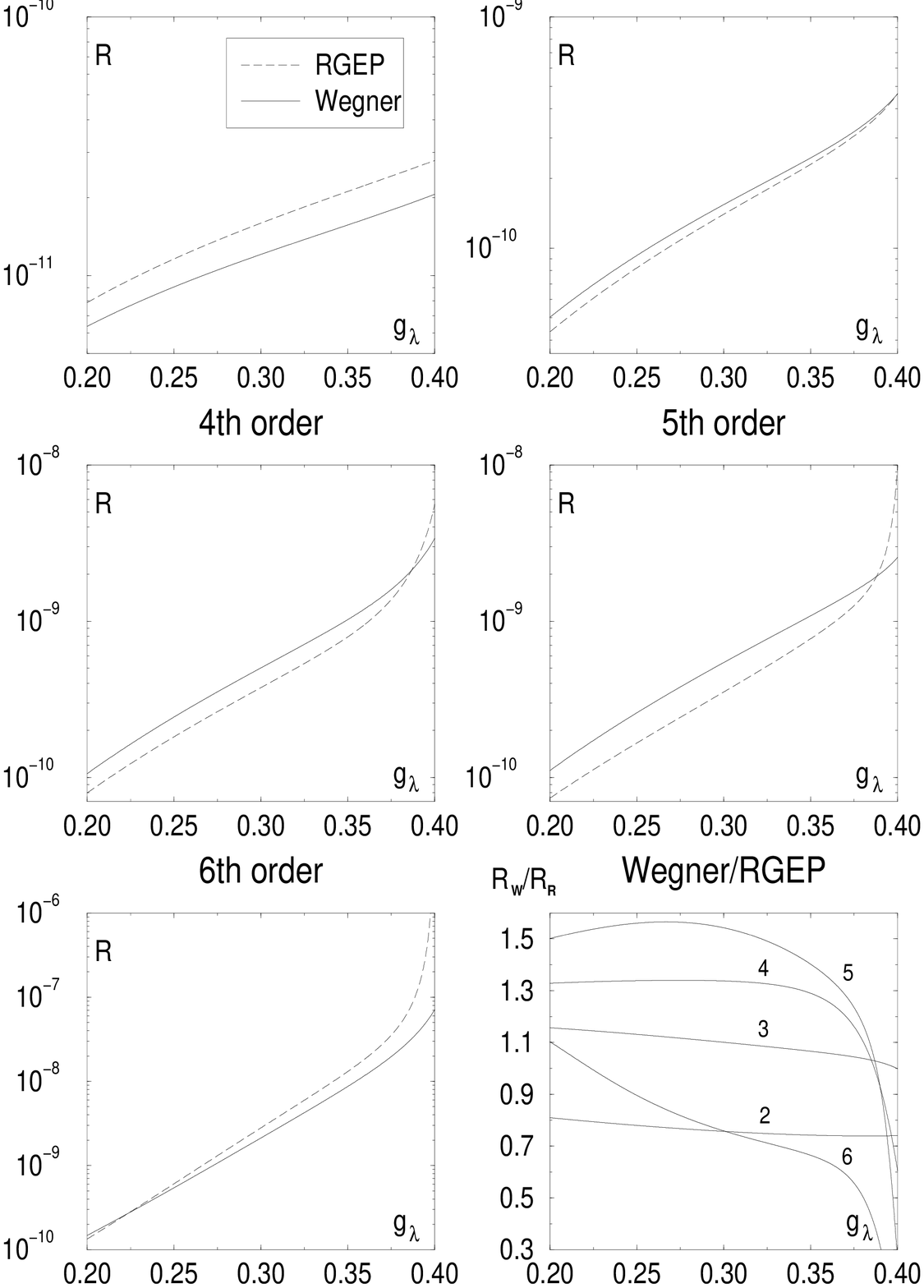}
\caption{Plots of the renormalizability measure $R(g_\lambda)$ from 
Eq. (\ref{eq:renormal}) for the RGEP procedure (dashed lines) and 
the altered Wegner equation (solid line). The last diagram displays
the ratios $R_W/R_R$ (with $R_W$ standing for the benchmark $R$, 
and $R_R$ for the $R$ in the RGEP procedure), and the integer labels
indicate the corresponding orders of the weak-coupling expansion.}
\label{renormalizowalnosc}
\end{figure}

The last question concerning accuracy of the RGEP method concerns 
renormalizability, which can be studied in analogy to \cite{optimization}.
However, one has to measure the sensitivity of the effective theory
at the scale $\lambda$ to the cutoff $\Lambda = b^N$ in a whole range 
of coupling constants, because the exact value of $g_\lambda$ is not 
known in the RGEP procedure. Using $g_{16}$ and $g_{20}$ from Section 
\ref{sec:model}, we calculate $W_\lambda$ starting from $H$ with the 
two different values of $\Lambda$ for $N = 16$ and $N = 20$, and we 
compare the resulting matrix elements of $W_\lambda$ at $\lambda = 2$ 
GeV. For $b=2$ the cutoff is changed by the factor 16, i.e. from 
about 65 TeV to about 1000 TeV. The divergence in the bare theory is 
logarithmic. The results could change at the rates implied by the 
change in the logarithm of $\Lambda$, i.e., about 25\%, but in the 
presence of a proper set of counterterms \cite{similarity} one expects 
no change to occur. Fig. \ref{renormalizowalnosc} shows the measure 
of changes in $W_\lambda$ that are obtained here with only one counterterm, 
which amounts to the change of the coupling constant from $g_{16}$ to 
$g_{20}$. The function plotted in \ref{renormalizowalnosc} is defined as
\begin{eqnarray}
R(g_\lambda)=\sum_{n,m=\tilde M}^{\tilde N}
\left[ \, \frac{W_{\lambda \, m n}(N=16)}
{W_{\lambda \, m n}(N=20)} \, - \, 1 \, \right]^2 \, .
\label{eq:renormal}
\end{eqnarray} 
The results change with the order of perturbation theory used for 
evaluation of $W_\lambda$. The sensitivity of results to the change 
of $\Lambda$ in 1st order is the same in RGEP as in the benchmark case
\cite{optimization}. Inclusion of higher orders, from 2 to 6, exhibits 
slight variations to the advantage of one or the other method, as 
shown in Fig. \ref{renormalizowalnosc} (the solid line for the benchmark 
and the dashed one for RGEP). The last diagram in Fig. \ref{renormalizowalnosc} 
shows the corresponding ratio  $R_W/R_R$, and it demonstrates that the 
renormalizability condition is satisfied with the same accuracy in both 
approaches. A closer comparison requires that the benchmark case (or $R_W$)
is taken for $g_\lambda = 0.2852$, and RGEP (or $R_R$) for some $g_\lambda 
\sim 0.3 \pm 0.005$ (see Appendix \ref{app:nd}), but these corrections are 
negligible at the current stage of the analysis. 
\section{Conclusion}
\label{sec:c}
The RGEP weak-coupling expansion achieves precision comparable with 
that of the best available benchmark method of the altered Wegner 
equation in the case of a simple matrix model for the Hamiltonian 
formulation of theories with asymptotic freedom and bound states. 
The few percent accuracy is reached by introducing a factor of $\phi$ 
in Eq. (\ref{factorphi}), which is similar in both methods. Since 
the RGEP procedure is designed for application to relativistic quantum 
field theory, the model test calculation provides a comprehensive 
outline of steps that can be repeated in realistic cases, especially 
in the theory of quarks and gluons. On the other hand, knowing that 
Wegner's equation is useful in condensed matter physics, one can 
expect that the altered Wegner equation and the RGEP procedure may 
also find applications in many-body theory. The model study indicates 
a possibility that the weak-coupling expansion may lead to a systematic
approximation scheme despite the growth of the coupling when the 
characteristic scale $\lambda$ of the effective theories is lowered. 
This is indicated in columns E and F in Figs. \ref{glambda} and \ref
{boundstate}, where one can see the convergence of the coupling constant 
to a stable value and the corresponding convergence of the bound-state 
energy to the exact result. One should also note that already second 
order calculations render effective window Hamiltonians $W_\lambda$
that can produce the bound-state eigenvalue with accuracy better 
than 10\%. 

The fact that both methods have similar accuracy suggests that they 
already show the range of calculational power that is available 
through a plain expansion in a single coupling constant such as 
the one defined in Eq. (\ref{defglambda}). In order to move beyond 
the 1\% accuracy level one has to achieve better understanding of 
the structure of $H_\lambda$. For example, there exist terms in 
$H_\lambda$ with specific dependence on the eigenvalues of $H_0$ and 
$\lambda$, like $(E_m - E_n)^2/\lambda^2$ in the model. One may 
hope to understand how the coefficients of these terms depend on a 
more suitably defined coupling constant than the one given Eq. (\ref
{defglambda}). It is conceivable that such understanding may further 
improve the weak-coupling expansion. Note also that none of the 
renormalization group universality features were so far explicitly 
employed in the plain expansion tested here. Such options for 
improvement may depend on a theory. One has to study specific 
theories using concrete versions of the RGEP procedure in order 
to identify the dominant terms and their universal behavior.

The accuracy study for the RGEP procedure shows also that one should
be able to carry out similar tests for any other approach to the 
bound-state problem in asymptotically free theories. For such a test 
to become possible, the approach in question would have to be understood 
sufficiently well to determine the steps that that approach implies for 
handling of the initial $H$ in the model. But such understanding is 
demanded of most formulations of relativistic quantum theories for 
fundamental reasons and the accuracy tests of similar kind can be 
consider a challenge for any scheme intended to solve the bound-state 
problem in theories with asymptotic freedom.
\appendix
\section{RGEP of 4th order}
\label{app:4th}
The RGEP weak-coupling expansion for terms of order $g^n$,
is written using terms of orders $k < n$,
\begin{eqnarray}
\mathcal{G}_n^{'} & = & 
\sum _{k=1}^{n-1}
\left[f\mathcal{G}_k,\left\{ \left(1-f\right)\mathcal{G}_{n-k}\right\}^{'}\right] \, .
\end{eqnarray}
If the interaction Hamiltonian $H_I = H - H_0$ is proportional to $g$,
the expansion in powers of $g$ is the same as the expansion in powers
of $H_I$. If $H_I$ contains a polynomial in $g$ with operator coefficients,
then the expansion in powers of $H_I$ is an intermediate step for obtaining 
an expansion for $H_\lambda$ in powers of $g$. Therefore, this appendix 
provides a generic set of coefficients for expansion of $\mathcal G_k(\lambda)$ 
in powers of $H_I$. The argument $\lambda$ is omitted in what follows. The 
procedure of rewriting the expansion in powers of the bare $g$ into the 
expansion in powers of $g_\lambda$ requires a definition of $g_\lambda$
in the structure of $H_\lambda$. This definition, in analogy to Eq. 
(\ref{defglambda}), provides then a series expression for $g_\lambda(g)$.
This expression is inverted and substituted into the formal expansion of 
$H_\lambda$ in powers of the bare $g$, producing the desired expansion
in powers of $g_\lambda$.  

The first four terms in the perturbative expansion are written as
\begin{eqnarray}
\left[\mathcal{G}_1\right]_{mn} & = & H_{mn} \, , \\
\left[\mathcal{G}_2\right]_{mn} & = & a_{m1n}H_{m1}H_{1n} \, , \\
\left[\mathcal{G}_3\right]_{mn} & = & b_{m12n}H_{m1}H_{12}H_{2n} \, , \\
\left[\mathcal{G}_4\right]_{mn} & = & c_{m123n}H_{m1}H_{12}H_{23}H_{3n} \, .
\end{eqnarray}
It is uderstood that the indices 1, 2, and 3, are summed over
the entire range available for them in the theory.

In order to write down expressions for the coefficients $a$, $b$, 
and $c$, we introduce a set of auxiliary symbols. Their meaning 
becomes sucessively self-evident when one decifers them in the 
order they are given here. The variables $s$ and $t$ have the 
meaning of the renormalization group parameter $1/\lambda^2$.
\begin{eqnarray}
E_{mn} & = & E_m -E_n \, , \\
G_{mn} & = & \varphi _{mn}\left(E_m-E_n\right) \, , \\
F_{mn} & = & \varphi _{mn}\left(E_m-E_n\right)^{2} \, , \\
G_{m1n} & = & G_{1m}+G_{1n} \, , \\
F_{m...lkn} & = & F_{m...lk}+F_{kn} \, , \\
A_{ml...kn}(t) & = & f_{ml}...f_{kn} \, , \\
A_{m...l+k...n}(t) & = & A_{m...l}(t)\, A_{k...n}(t) \, , \\
B_{m...k}(s) & = & \int _{0}^{s}dtA_{m...k}(t) \, .
\end{eqnarray}
In this notation one obtains
\begin{eqnarray}
a_{m1n} & =G_{m1n}\,  & B_{m1n} \, .
\end{eqnarray}
\begin{eqnarray}
b_{m12n} & = & \frac{G_{m12}}{E_{m2}}\left[B_{m12n+m2}-B_{m12n}\right] \nonumber \\
 & + & \frac{G_{n12}}{E_{n2}}\left[B_{m21n+n2}-B_{m21n}\right] \nonumber \\
 & + & \frac{G_{m1n}\, G_{m12}}{F_{m12}}\left[B_{m2n+m12}-B_{m2n}\right] \nonumber \\
 & + & \frac{G_{m1n}\, G_{n12}}{F_{n12}}\left[B_{m2n+n12}-B_{m2n}\right]
\end{eqnarray}
\begin{widetext}
\begin{eqnarray}
c_{m123n} & = & \frac{G_{23m}}{E_{m1}\, E_{m2}}\left[B_{m321n}-
B_{m321n+1m}-B_{m321n+2m}+B_{m321n+1m2}\right] \nonumber \\
 & + & \frac{G_{23n}}{E_{n1}\, E_{n2}}\left[B_{m123n}-B_{m123n+1n}-
B_{m123n+2n}+B_{m123n+1n2}\right] \nonumber \\
 & + & \frac{G_{132}}{E_{12}\, E_{m1}}\left[B_{m231n}-B_{m231n+21}-
B_{m231n+m1}+B_{m231n+m12}\right] \nonumber \\
 & + & \frac{G_{132}}{E_{12}\, E_{n1}}\left[B_{m132n}-B_{m132n+21}-
B_{m132n+n1}+B_{m132n+n12}\right] \nonumber \\
 & + & \frac{G_{12m}\, G_{23m}}{F_{23m}\, E_{m1}}\left[B_{m21n}-
B_{m21n+m1}-B_{m21n+m32}+B_{m21n+1m32}\right] \nonumber \\
 & + & \frac{G_{12n}\, G_{23n}}{F_{23n}\, E_{n1}}\left[B_{m12n}-
B_{m12n+n1}-B_{m12n+n32}+B_{m12n+1n32}\right] \nonumber \\
 & + & \frac{G_{12m}\, G_{13n}}{F_{13n}\, E_{m1}}\left[B_{m21n}-
B_{m21n+m1}-B_{m21n+n31}+B_{m21n+m13n}\right] \nonumber \\
 & + & \frac{G_{12n}\, G_{13m}}{F_{13m}\, E_{n1}}\left[B_{m12n}-
B_{m12n+n1}-B_{m12n+m31}+B_{m12n+m31n}\right] \nonumber \\
 & + & \frac{G_{12m}\, G_{132}}{F_{132}\, E_{m1}}\left[B_{m21n}-
B_{m21n+m1}-B_{m21n+132}+B_{m21n+m132}\right] \nonumber \\
 & + & \frac{G_{12n}\, G_{132}}{F_{132}\, E_{n1}}\left[B_{m12n}-
B_{m12n+n1}-B_{m12n+132}+B_{m12n+n132}\right] \nonumber \\
 & + & \frac{G_{m1n}\, G_{23m}}{E_{m2}}\left[\frac{B_{m1n}-
B_{m1n+m321}}{F_{m321}}-\frac{B_{m1n}-B_{m1n+m321+m2}}{F_{m321}+F_{2m}}\right] \nonumber \\
 & + & \frac{G_{m1n}\, G_{23n}}{E_{n2}}\left[\frac{B_{m1n}-
B_{m1n+n321}}{F_{n321}}-\frac{B_{m1n}-B_{m1n+n321+n2}}{F_{n321}+F_{2n}}\right] \nonumber \\
 & + & \frac{G_{m1n}\, G_{132}}{E_{12}}\left[\frac{B_{m1n}-
B_{m1n+m231}}{F_{m231}}-\frac{B_{m1n}-B_{m1n+m2132}}{F_{m231}+F_{21}}\right] \nonumber \\
 & + & \frac{G_{m1n}\, G_{132}}{E_{12}}\left[\frac{B_{m1n}-
B_{m1n+n231}}{F_{n231}}-\frac{B_{m1n}-B_{m1n+n2132}}{F_{n231}+F_{21}}\right] \nonumber \\
 & + & \frac{G_{m1n}\, G_{12m}\, G_{23m}}{F_{23m}}\left[\frac{B_{m1n}-
B_{m1n+m21}}{F_{m21}}-\frac{B_{m1n}-B_{m1n+m321+m2}}{F_{m21}+F_{m32}}\right] \nonumber \\
 & + & \frac{G_{m1n}\, G_{12n}\, G_{23n}\, }{F_{23n}}\left[\frac{B_{m1n}-
B_{m1n+n21}}{F_{n21}}-\frac{B_{m1n}-B_{m1n+n321+n2}}{F_{n21}+F_{n32}}\right] \nonumber \\
 & + & \frac{G_{m1n}\, G_{12m}\, G_{132}}{F_{132}}\left[\frac{B_{m1n}-
B_{m1n+m21}}{F_{m21}}-\frac{B_{m1n}-B_{m1n+m2312}}{F_{m21}+F_{231}}\right] \nonumber \\
 & + & \frac{G_{m1n}\, G_{12n}\, G_{132}\, }{F_{132}}\left[\frac{B_{m1n}-
B_{m1n+n21}}{F_{n21}}-\frac{B_{m1n}-B_{m1n+n2312}}{F_{n21}+F_{231}}\right] \nonumber \\
 & + & \frac{G_{m1n}\, G_{12n}\, G_{13m}}{F_{12n}\, F_{13m}}\left[B_{m1n}-
B_{m1n+n21}-B_{m1n+m31}+B_{m1n+m312n}\right]
\end{eqnarray}
\end{widetext}
In relativistic applications, these formulae need a replacement of the 
differences of energies like $E_m - E_n$ by the changes of invariant 
masses in the interaction vertices, and multiplication of $G_{mn}$ by
the parent momenta $P^+_{mn}$ in the vertices, see e.g. \cite{g_QCD, ho}.

\newpage
\clearpage

\begin{figure}
\includegraphics[width=84mm,height=210mm]{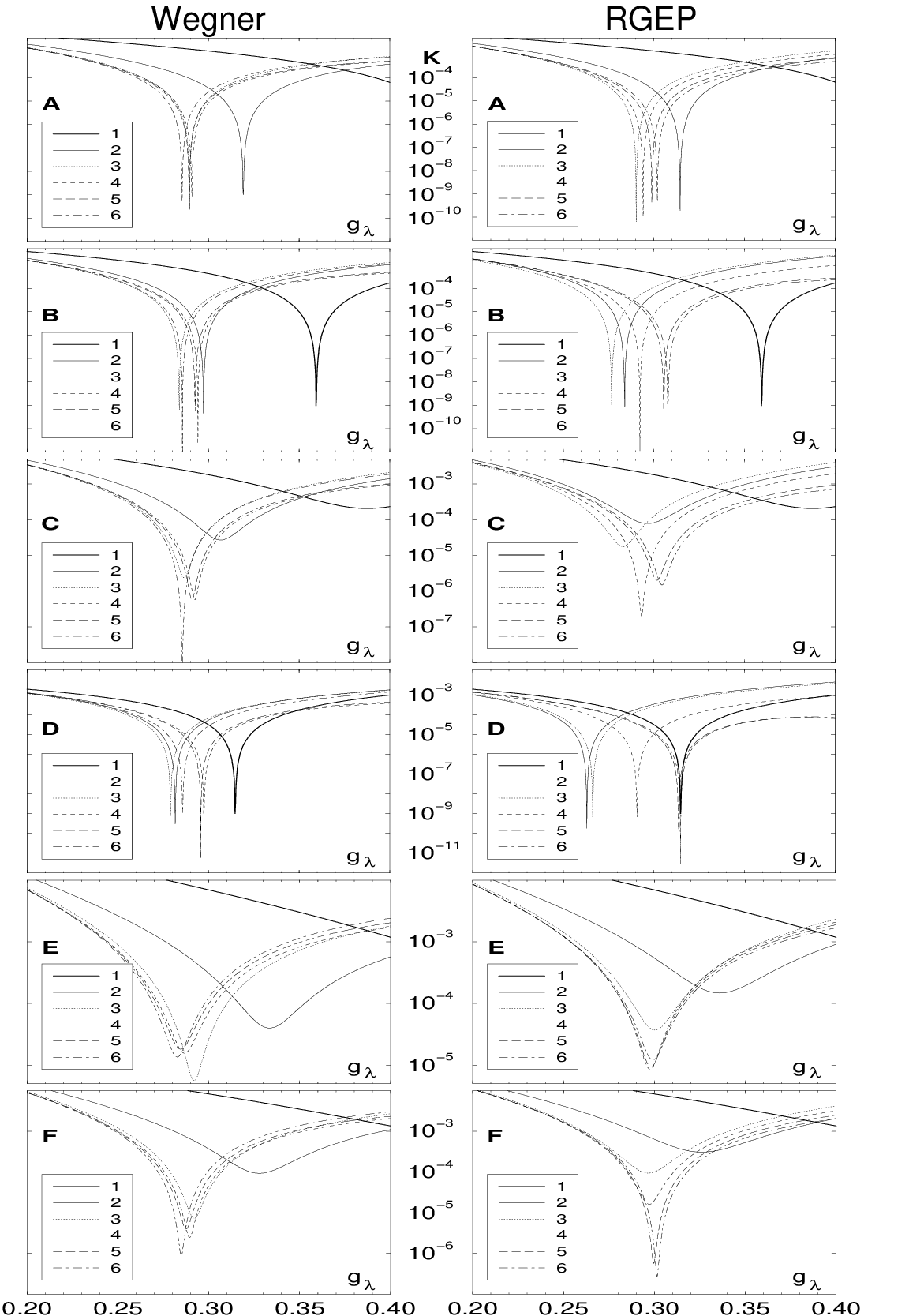}
\caption{The functions $K(g_\lambda)$ from Eqs. (\ref{caseA}) to (\ref{caseF})
plotted for Wegner and RGEP cases, $\lambda = 2$ GeV. In all cases a preferred 
value of $g_\lambda$ is clearly identified. These are given in Table I.}
\label{fitowanie}
\end{figure}
\begin{table}
\label{tab:1}
\begin{tabular}{|c||rl||rl|}
\hline 
\multicolumn{1}{c}{               }&
\multicolumn{2}{c}{\textbf{Wegner}}&
\multicolumn{2}{c}{\textbf{RGEP}}\\
\hline
\hline 
$ k          $ &
$g_\lambda   $ &
$|E_k|       $ &
$g_\lambda   $ &
$|E_k|       $ \\
\hline
\multicolumn{5}{c}{\textbf{A}}\\
\hline
1 &  0.43340  & 0.830955  & 0.43340  & 0.830955 \\
2 &  0.31900  & 0.984874  & 0.31460  & 0.826301 \\ 
3 &  0.28985  & 1.006771  & 0.29040  & 0.864725 \\
4 &  0.29095  & 1.045453  & 0.29370  & 0.911430 \\
5 &  0.28930  & 1.032121  & 0.29865  & 0.958206 \\
6 &  0.28545  & 1.005530  & 0.30195  & 0.987875 \\
\hline 
\multicolumn{5}{c}{\textbf{B}}\\
\hline
1 &  0.35915  & 0.539380  & 0.35915  & 0.539380 \\
2 &  0.29700  & 0.807142  & 0.28380  & 0.619971 \\
3 &  0.28380  & 0.943143  & 0.27665  & 0.746829 \\
4 &  0.29425  & 1.082537  & 0.29205  & 0.896377 \\
5 &  0.29260  & 1.069262  & 0.30525  & 1.020620 \\
6 &  0.28545  & 1.005530  & 0.29205  & 0.896377 \\
\hline 
\multicolumn{5}{c}{\textbf{C}}\\
\hline
1 &  0.38720  & 0.644935  & 0.38720  & 0.644935 \\
2 &  0.30690  & 0.885308  & 0.29645  & 0.701781 \\
3 &  0.28655  & 0.971815  & 0.28270  & 0.797715 \\
4 &  0.29260  & 1.063916  & 0.29315  & 0.906400 \\
5 &  0.29095  & 1.050609  & 0.30140  & 0.984000 \\
6 &  0.28545  & 1.005530  & 0.30415  & 1.008693 \\
\hline 
\multicolumn{5}{c}{\bf{D}}\\
\hline
1 &  0.31460  & 0.385414  & 0.31460  & 0.385414 \\
2 &  0.28160  & 0.691679  & 0.26290  & 0.494306 \\
3 &  0.27885  & 0.892589  & 0.26620  & 0.662643 \\
4 &  0.29755  & 1.120255  & 0.29040  & 0.881438 \\
5 &  0.29590  & 1.107059  & 0.31350  & 1.101064 \\
6 &  0.28545  & 1.005530  & 0.31460  & 1.110135 \\
\hline 
\multicolumn{5}{c}{\bf{E}}\\
\hline
1 &  0.48345  & 1.046788  & 0.48345  & 1.046788 \\
2 &  0.33330  & 1.108054  & 0.33605  & 0.983708 \\
3 &  0.29205  & 1.030408  & 0.30030  & 0.954532 \\
4 &  0.28655  & 0.996998  & 0.29755  & 0.946995 \\
5 &  0.28435  & 0.977641  & 0.29810  & 0.953083 \\
6 &  0.28270  & 0.974623  & 0.29920  & 0.962119 \\
\hline 
\multicolumn{5}{c}{\bf{F}}\\
\hline
1 &  0.45760  & 0.933635  & 0.45760  & 0.933635 \\
2 &  0.32780  & 1.059985  & 0.32615  & 0.909722 \\
3 &  0.29205  & 1.030408  & 0.29645  & 0.919122 \\
4 &  0.28930  & 1.027150  & 0.29700  & 0.941877 \\
5 &  0.28765  & 1.013796  & 0.29975  & 0.968487 \\
6 &  0.28490  & 0.999307  & 0.30140  & 0.982700 \\
\hline
\end{tabular}
\caption{The values of the couplings $g_\lambda$, which correspond
to the minima of curves in Fig. \ref{fitowanie} (the last digit is
given with error margin of 5), and the corresponding results (obtained 
with the coupling constants given in this table) for the bound state 
energy $|E_k|$ (the exact value is 1), obtained by diagonalization of 
$W_\lambda$ with $\lambda = 2$ GeV, that was calculated in six 
successive orders of the weak-coupling expansion, $k$.}
\end{table}
\clearpage

\section{Numerical details}
\label{app:nd}
Numerical integration of Eqs. (\ref{eq:rgrwegner}) and (\ref{master})
was performed using fourth-order Runge-Kutta procedure and an automated
algorithm for generating expressions of order $n$ from expressions of
orders $k < n$. The numerically calculated matrix elements $H_{\lambda 
\, mn}$ were checked using analytic formula from Appendix \ref{app:4th}. 

The fits of the coupling constants $g_\lambda$ that were used in the 
perturbative evaluation of the windows $W_\lambda$, were obtained using 
the functions $K(g_\lambda)$ from Eqs. (\ref{eq:fit1}) and (\ref{eq:fit2})
that are plotted in Fig. \ref{fitowanie}. The corresponding numbers 
for the coupling constants are given in Table I.

\end{document}